\documentclass[twocolumn,noshowpacs,amsfonts,amssymb,amsmath,letterpaper,floatfix]{revtex4}

\usepackage{graphicx}  
\usepackage{setspace}

\begin{document}   
\newlength{\GraphicsWidth}
\setlength{\GraphicsWidth}{8cm}     

\newcommand\comment[1]{\textsc{{#1}}}
\newcommand{\Label}[1]{\label{#1}}  
\renewcommand{\r}{\mathbf{r}}
\newcommand{\be}{\begin{equation}}
\newcommand{\ee}{\end{equation}}

\newcommand{\Man}{{\text{Manning}}}
\newcommand{\hr}{{\tilde{r}}}
\newcommand{\ha}{{\tilde{a}}}
\newcommand{\XXXX}{{\bf XXXX}}
\newcommand{\eff}{{\text{eff}}}
\newcommand{\sat}{{\text{sat}}}

\title{%
Preferential interaction coefficient for nucleic acids and other cylindrical poly-ions
}
\author{Emmanuel Trizac}
\email{trizac@lptms.u-psud.fr}
\affiliation{CNRS; Univ. Paris Sud, UMR8626, LPTMS, ORSAY CEDEX, F-91405}
\affiliation{Center for Theoretical Biological Physics, UC San Diego,                    
       9500 Gilman Drive MC 0374 - La Jolla, CA  92093-0374, USA}
\author{Gabriel T\'ellez}
\email{gtellez@uniandes.edu.co}
\affiliation{Departamento de F\'{\i}sica, Universidad de Los Andes,
Apartado A\'ereo~4976, Bogot\'a, Colombia}

\begin{abstract}                              
The thermodynamics of nucleic acid processes is heavily affected 
by the electric double-layer of micro-ions around the polyions.
We focus here on the 
Coulombic contribution to the salt-polyelectrolyte preferential interaction
(Donnan) coefficient and we report extremely accurate analytical
expressions valid in the range of low salt
concentration (when polyion radius is smaller than the
Debye length). The analysis is performed at Poisson-Boltzmann
level, in cylindrical geometry, with emphasis on highly charged poly-ions 
(beyond ``counter-ion condensation''). 
The results hold for any electrolyte of the form $z_-$:$z_+$.
We also obtain a remarkably accurate expression for the electric potential 
in the vicinity of the poly-ion.
\end{abstract}


\maketitle



Coulombic interactions between salt and poly-anions 
play a key role in the equilibrium and kinetics of nucleic
acid processes \cite{Anderson}. A convenient quantity
quantifying such interactions and allowing for the analysis
and interpretation of their thermodynamics consequences, is
the so called preferential interaction coefficient. Several
definitions have been proposed and their interrelation studied,
see e.g. \cite{Eisenberg,Sch,Tim}. In the present work, 
they are defined as the integrated deficit (with respect to bulk
conditions) of co-ions concentration around
a rod-like poly-ion. Our goal is to provide analytical 
expressions describing the effect of salt concentration and poly-ion 
structural parameters on the preferential interaction coefficient,
for a broad class of asymmetric electrolytes.
For symmetric electrolytes, it will be shown that our formulas improve 
upon existing analytical results. For other asymmetries, they seem to have
no counterpart in the literature.
Our analysis holds  for highly (i.e. beyond counter-ion condensation
\cite{Manning,Oosawa}) 
and uniformly charged cylindrical poly-ions, and is explicitly
limited to the low
salt regime (i.e. when the poly-ion radius $a$ is smaller than 
the Debye length $1/\kappa$). These conditions are  most relevant  
for RNA or DNA in their single, double, or triple strand forms.

As in several previous approaches \cite{Sharp,Ni,Shkel,Taubes}, 
we adopt the mean-field framework
of Poisson-Boltzmann equation, in a homogeneous dielectric 
background of permittivity $\varepsilon$. 
The same starting point has proven relevant for related 
structural physical chemistry studies of nucleic acids 
\cite{Gueronbis}.
In a $z_-$:$z_+$ electrolyte,
the dimensionless electrostatic potential $\phi = e \varphi/kT$
(with $e>0$ the elementary charge and $kT$ thermal energy) 
then obeys the equation \cite{Levin}
\be
\frac{1}{r}\frac{d}{dr}\left(r \frac{d \phi}{d r}\right) \,=\, 
\frac{\kappa^2}{z_+ + z_-} \,\left[ e^{z_- \phi}-
e^{-z_+ \phi} 
\right],
\Label{eq:PB}
\ee
where $r$ is the radial distance to the rod axis. 
The valencies $z_+$ and $z_-$ of salt ions are both taken positive.
Denoting derivative with a prime, 
the boundary conditions read $r \phi'(r) = 2\xi >0$ at the polyion
radius ($r=a$) 
and $\phi \to 0$ for $r\to \infty$. The latter condition
expresses the infinite dilution of poly-ion limit and ensures that the
whole system is electrically neutral, since it (indirectly) implies that 
$r \phi' \to 0$ for $r \to \infty$. 
We consider a negatively charged poly-anion for which $\phi<0$ and the 
line charge density reads $\lambda = -e \xi /\ell_B<0$,
where $\ell_B = e^2/(\varepsilon kT)$ denotes the Bjerrum length
(0.71 nm in water at room temperature). Finally, the Debye length
is defined from the bulk ionic densities $n_+^\infty$ and $n_-^\infty$
through $\kappa^2 = 4 \pi \ell_B (z_+^2 n_+^\infty + z_-^2 n_-^\infty)$.

The Coulombic contribution to the anionic preferential interaction coefficient is defined as
\cite{Sharp,Ni,Shkel,Taubes,rque10}
\be
\Gamma \,=\, \kappa^2 \,\int_a^\infty (e^{z_- \phi}-1) \, r dr,
\Label{eq:Gamma}
\ee
while its cationic counterpart follows from electro-neutrality.
This quantity --which provides a measure of the Donnan effect \cite{Donnan}--
can be expressed in closed form as a function 
of the electrostatic potential, see Appendix \ref{app:A}.
As can be seen in (\ref{eq:a3}) and (\ref{eq:a4}), $\Gamma$
depends exponentially on the surface potential $\phi_0$, 
so that deriving a precise analytical expression is a challenging task.
Furthermore, we are interested here in the limit $\kappa a <1$
(including the regime $\kappa a \ll 1$) which is analytically more difficult
than the opposite high salt situation where to leading order,
the charged rod behaves as an infinite plane, and curvature
corrections can be perturbatively included \cite{Shkelhighsalt,JPA2003,PRE2004}.

We will proceed in two steps. Focusing first on the 
surface potential $\phi_0= \phi(a)$, we make use of recent results 
\cite{TT} that have been obtained from a mapping of Eq. (\ref{eq:PB})
onto a Painlev\'e type III problem \cite{McCoy,McCaskill,Tracy}. The exact expressions
thereby derived only hold for 1:1, 1:2 and 2:1 electrolytes, but may be written 
in a way that is electrolyte independent. This remarkable feature
is specific to the short distance behaviour of $\phi$ and has been 
overlooked so far, since not only short distance but also large distance
properties have been studied \cite{TT}. We are then led to conjecture that
the corresponding expression holds for {\em any} binary electrolyte
$z_-$:$z_+$, and we explicitly check the relevance of our assumption
on several specific examples. 

Technical details are deferred to the appendices. It is in particular concluded
in Appendix \ref{app:B} that the surface potential may be written 
\be
e^{-z_+ \phi_0} \,\simeq\, \frac{2 (z_++z_-)}{z_+ (\kappa a)^2} \,
\left[(z_+ \xi-1)^2 +\widetilde\mu^2
\right]
\Label{eq:potsurf}
\ee
where
\be
\widetilde\mu \, \simeq\, \frac{-\pi}{\log(\kappa a) +{\cal C} -(z_+\xi-1)^{-1}} .
\Label{eq:mutilde}
\ee
Expression (\ref{eq:mutilde}) is valid for $\kappa a<1$ and $z_+ \xi >1$ [in fact 
$z_+\xi>1+{\cal O}(1/|\log \kappa a|)$]. These conditions are easily fulfilled
for nucleic acids. 
The ``constant''  $\cal C$ appearing in (\ref{eq:potsurf})
depends smoothly on the ratio $z_+/z_-$ but is otherwise salt and
charge independent. We report in Table \ref{table:1} 
its values for several electrolyte asymmetries. 
The decrease (in absolute value) of $\cal C$ when $z_+/z_-$
increases is a signature of more efficient (non-linear) screening with
counter-ions of higher valencies.

\begin{center}
\begin{table}
\begin{tabular}{c||ccccccc}
$z_+/z_-$&  1/10  &  1/3    &   1/2    &     1       &      2      &    3   &    10    \\
\hline\hline
${\cal C}$&   -2.51& -1.94 &   -1.763 & -1.502&-1.301& -1.21 & -1.06
\end{tabular}
\caption{Values of ${\cal C}$ appearing in Eq. (\ref{eq:mutilde}) as a function 
of electrolyte asymmetries. For $z_+/z_- = 1$, $1/2$ and 2,
${\cal C}$ is known analytically from the results of \cite{TT}. The 
corresponding values are
recalled in Appendix \ref{app:B}.
For other values  of $z_+/z_-$, $\cal C$ has been determined
numerically, see in particular Fig. \ref{fig:Q1} of Appendix \ref{app:B}.}
\label{table:1}
\end{table}
\end{center}

\begin{figure}[hb]
\vskip 4.5mm
\includegraphics[height=6cm,angle=0]{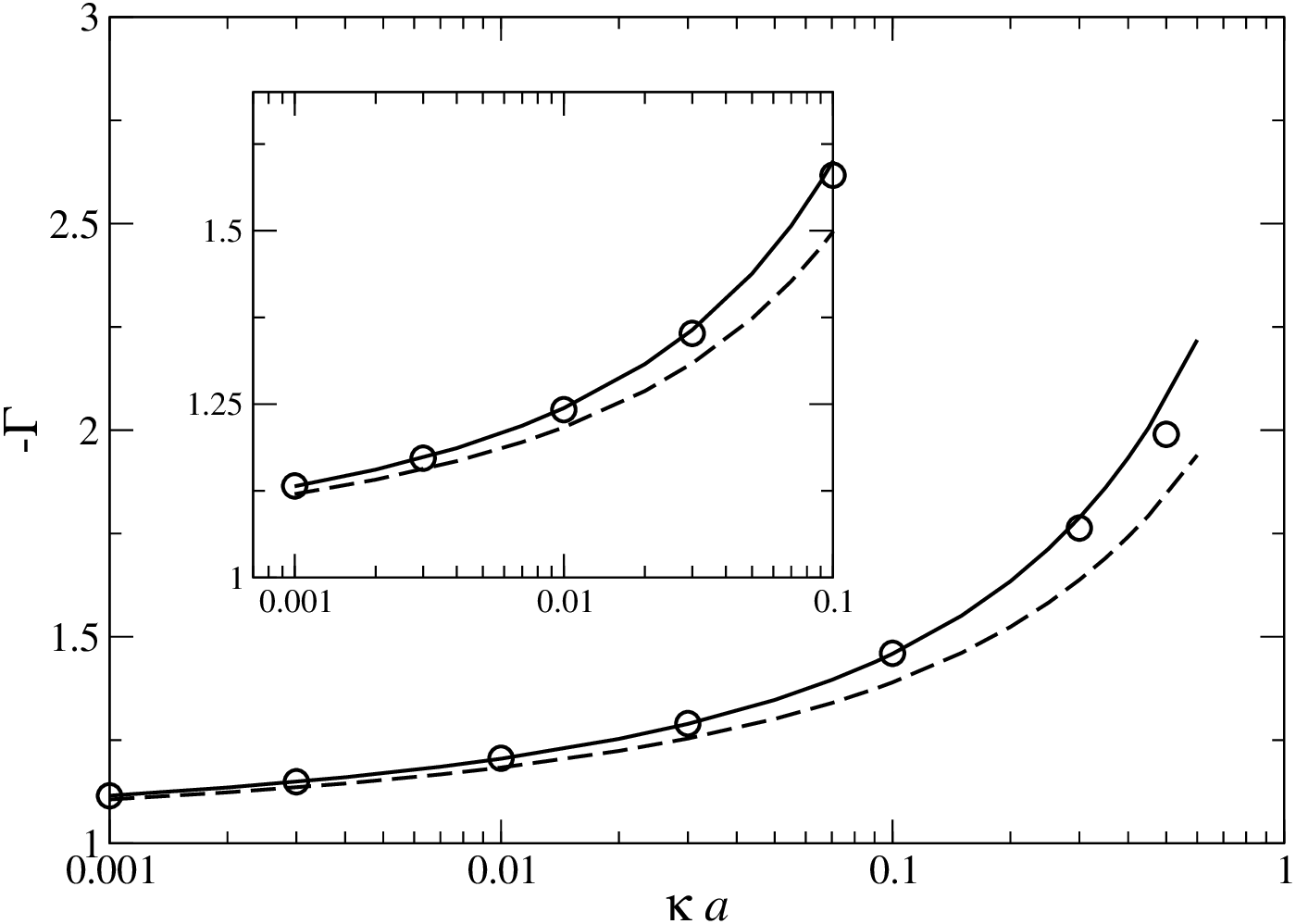}
\caption{Preferential interaction coefficient for a 1:1 salt. The main graph
corresponds to ss-RNA with reduced line charge $\xi=2.2$ while the
inset is for ds-RNA ($\xi=5$). The circles correspond to the
value of (\ref{eq:Gamma}) following from the numerical solution 
of Eq. (\ref{eq:PB}). The prediction of Eq. (\ref{eq:Gammaapprox})
with $\widetilde\mu$ given by (\ref{eq:mutilde}) and ${\cal C}\simeq -1.502$,
shown with the continuous curve, is compared to that of Ref. 
\cite{Shkel}, shown with the dashed line. As in all other figures,
the opposite of $\Gamma$ is displayed, to consider a positive quantity.
\Label{fig:Gamma11}
}
\end{figure}

From Eq. (\ref{eq:potsurf}) and the results of Appendix \ref{app:B},
our approximation for $\Gamma$ takes a simple form
\be
\Gamma \,\simeq \, -\frac{ z_-}{z_+} \,(1+\widetilde\mu^2).
\label{eq:Gammaapprox}
\ee
This expression is tested  in Figures \ref{fig:Gamma11} and \ref{fig:Gamma}
against the ``true'' numerical results that serve
as a benchmark.
In Fig. \ref{fig:Gamma11} which corresponds to a monovalent
salt (or more generally a $z$:$z$ electrolyte), we also
show the prediction of Ref. \cite{Shkel},
which is, to our knowledge, the most accurate existing formula
for a 1:1 salt. For the technical reasons discussed in Appendix \ref{app:B},
and that are evidenced in Figure \ref{fig:Q1}, our expression 
improves that of Shkel, Tsodikov and Record \cite{Shkel},
particularly at lower salt content.
For 1:2 and 2:1 salts, we expect Eq. (\ref{eq:Gammaapprox})
to be also accurate, since it is based on exact expansions.
The situation of other salt asymmetries is more conjectural
(see Appendix \ref{app:B}), but Eq. (\ref{eq:Gammaapprox}) is 
nevertheless in remarkable 
agreement with the full solution of Eq. (\ref{eq:PB}),
see Fig. \ref{fig:Gamma}. To be specific, in both Figures
\ref{fig:Gamma11} and \ref{fig:Gamma}, the relative accuracy 
of our approximation is better than 0.2\% for $\kappa a=10^{-2}$
(for both ss and ds RNA parameters). At $\kappa a=0.1$,
the accuracy is on the order of 1\%.

\begin{figure}[htb]
\vskip 4.5mm
\includegraphics[height=6cm,angle=0]{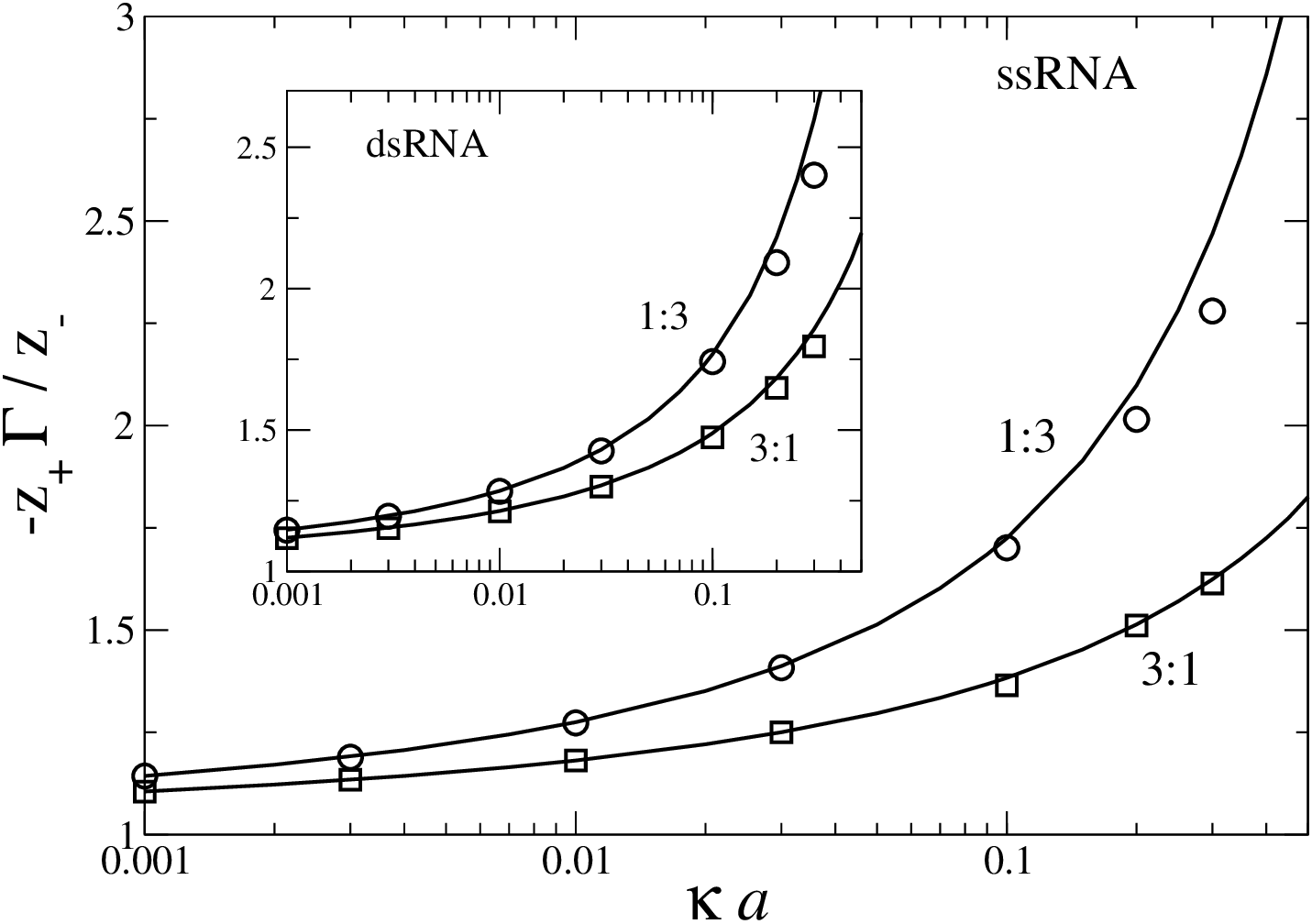}
\caption{Same as Figure \ref{fig:Gamma11} for a 1:3 and a 3:1 
electrolyte. From Table \ref{table:1}, we have ${\cal C} \simeq -1.21$
in the 1:3 case and conversely ${\cal C}\simeq -1.94$ in the 3:1 case.
The symbols correspond to the numerical solution of Eq. 
(\ref{eq:PB}) and the continuous curves show the results of
Eq. (\ref{eq:Gammaapprox}) with again $\widetilde\mu$
given by (\ref{eq:mutilde}).
\Label{fig:Gamma}
}
\end{figure}

\begin{figure}[htb]
\vskip 4.5mm
\includegraphics[height=6cm,angle=0]{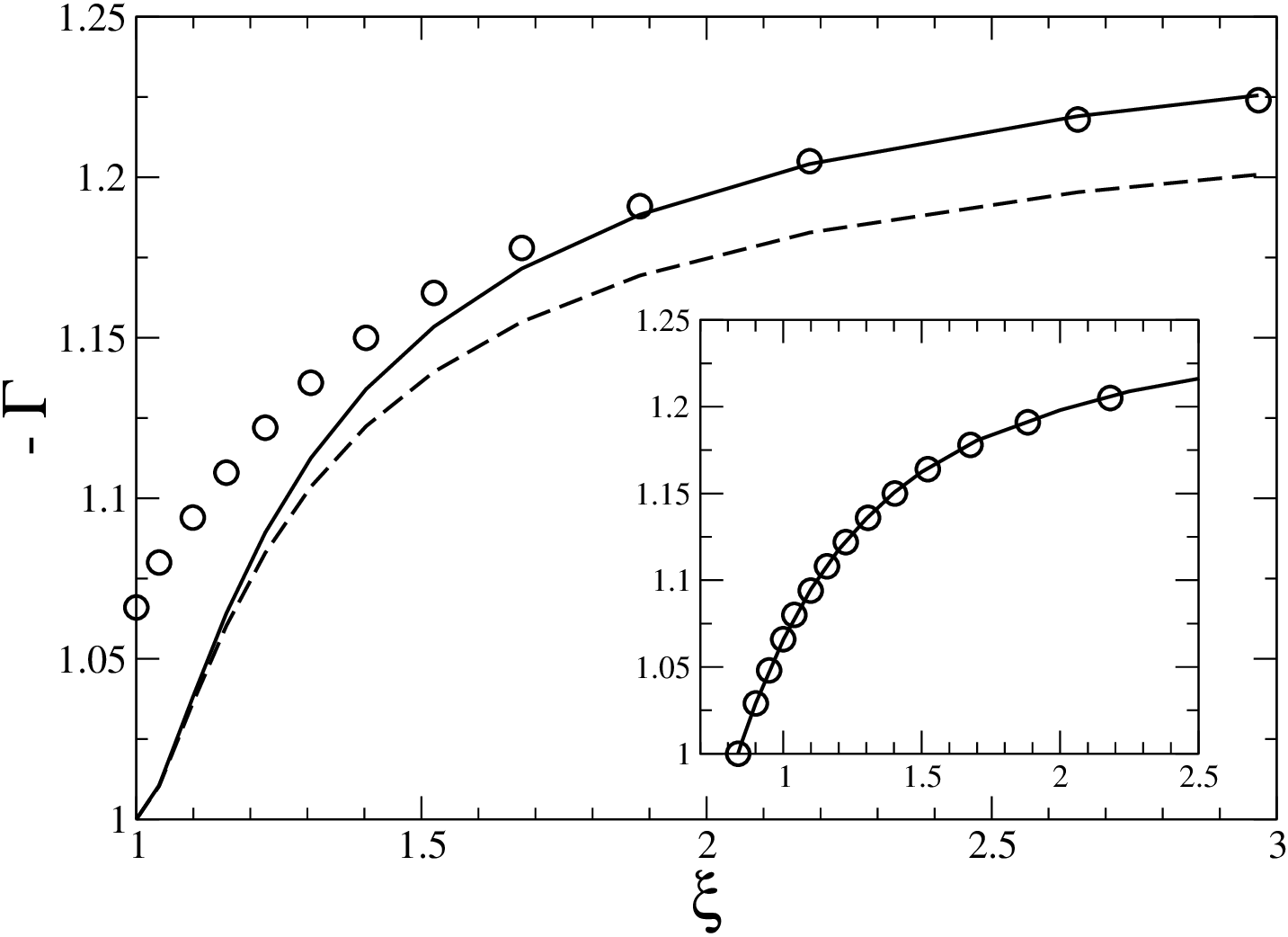}
\caption{Preferential interaction coefficient
for a 1:1 salt (hence ${\cal C}\simeq -1.502$)
and $\kappa a=10^{-2}$. The circles show the numerical 
solution of PB theory (\ref{eq:PB}), the continuous curve
is for (\ref{eq:Gammaapprox}) with (\ref{eq:mutilde}) and 
the dashed line is the prediction of Ref. \cite{Shkel}. 
Although approximation (\ref{eq:mutilde}) breaks down 
at low $\xi$, the inset shows that $\widetilde\mu$ following from the
solution of Eq. (\ref{eq:mutantilde}) gives through (\ref{eq:Gammaapprox})
a $\Gamma$ (continuous curve), that is in excellent agreement with
the ``exact one'', shown with circles as in the main graph. 
\Label{fig:Gamkap0.01}
}
\end{figure}

As illustrated in Fig. \ref{fig:Gamkap0.01},  
approximation (\ref{eq:mutilde}) assumes that $z_+ \xi>1$.
The corresponding expression for $\Gamma$ therefore
breaks down when $\xi$ is too low.
More general expressions, still for $\kappa a<1$,
may be found in appendix \ref{app:C}. 
The inset of Fig. \ref{fig:Gamkap0.01} offers an
illustration and shows that the limitations of approximation
(\ref{eq:mutilde}) may be circumvented at little cost,
providing a quasi-exact value for $\Gamma$. 
Moreover,  it is shown in this appendix that for $z_+ \xi=1$,
$\widetilde\mu$ reads
\be
\widetilde\mu  \, \simeq\, \frac{-\pi/2}{\log(\kappa a) +{\cal C}} .
\label{eq:muxi1}
\ee
On the other hand, Eq. (\ref{eq:potsurf}) still holds.
The corresponding $\Gamma$ is shown in Fig. 
\ref{fig:Gamxi1}.

\begin{figure}[htb]
\vskip 4.5mm
\includegraphics[height=6cm,angle=0]{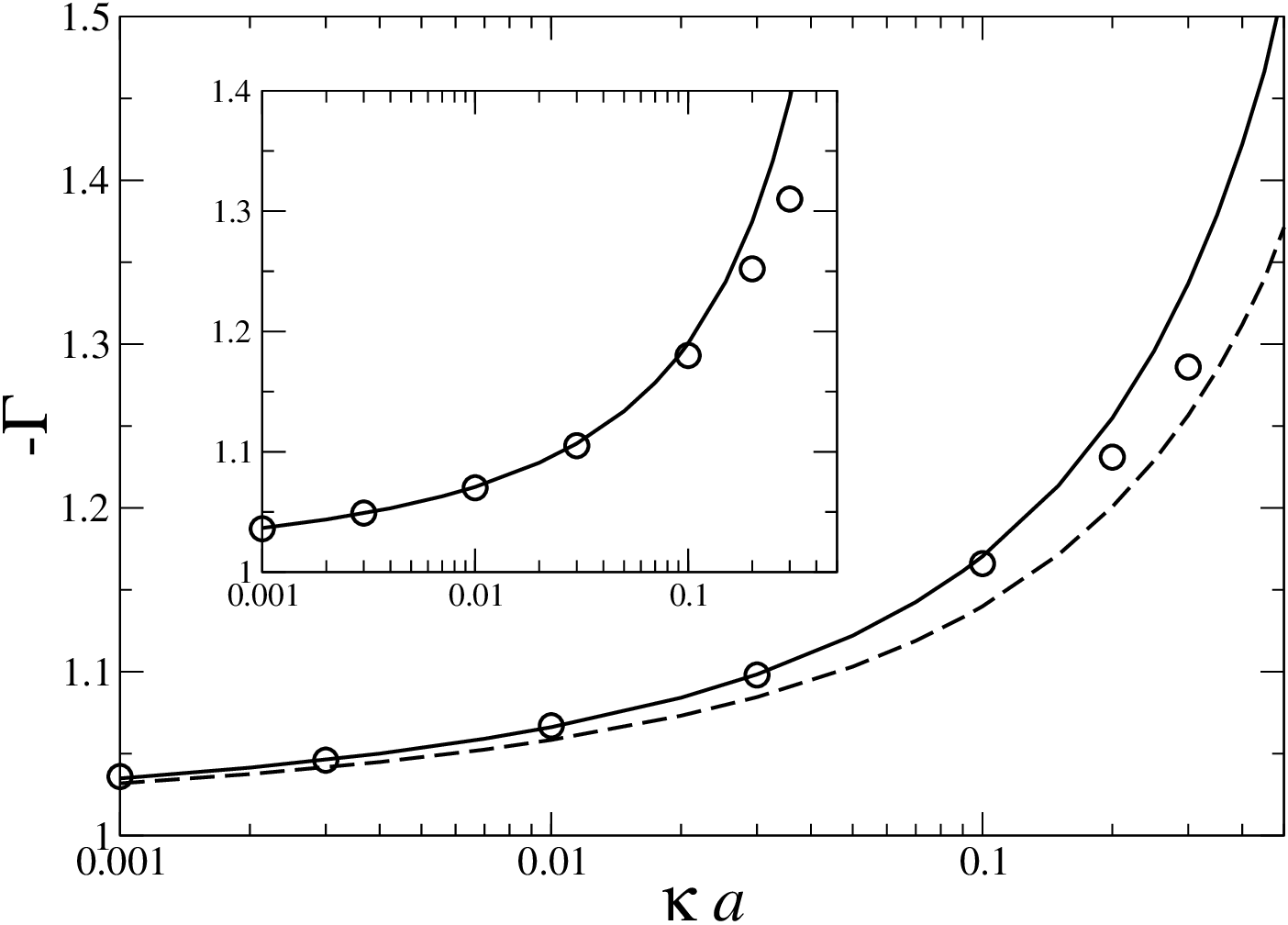}
\caption{Same as Fig. \ref{fig:Gamma11} for $\xi=1$ and
$z_+/z_-=1$. The same quantities are shown: 
our prediction for $\Gamma$ [Eqs. (\ref{eq:Gammaapprox})
and (\ref{eq:muxi1}) with ${\cal C}\simeq -1.502$] 
is compared to that of Ref. \cite{Shkel}. 
The inset shows $-z_+\Gamma/z_-$ for a 1:2 salt such as MgCl$_2$
where $\cal C$ takes the value -1.301.
Circles : numerical data; curve : our prediction. 
\Label{fig:Gamxi1}
}
\end{figure}

We provide in Appendix \ref{app:C} a general expression of the
short scale (i.e valid up to $\kappa r \sim 1$)
radial dependence of the electric potential,
see Eq. (\ref{eq:phigen}).
The bare charge should not be too low [more precisely, one must have
$\xi>\xi_c$ with $\xi_c$ given by Eq. (\ref{eq:xicapp})],
and $\widetilde\mu$ --which encodes the dependence on $\xi$-- 
follows from solving Eq.~(\ref{eq:mutantilde}).
In general, the corresponding solution should be found 
numerically. However, one can show {\em a)} that $\widetilde\mu$
vanishes for $\xi=\xi_c$, {\em b)} that $\widetilde\mu$ takes the value
(\ref{eq:muxi1}) when $z_+\xi=1$ and {\em c)} that 
$\widetilde\mu$ is given by (\ref{eq:mutilde}) when $z_+\xi$
exceeds unity by a small and salt dependent amount.
In practice, for DNA and RNA, we have $\xi>2$ and
Eq. (\ref{eq:mutilde}) provides excellent results
whenever $\kappa a<0.1$. To illustrate this, we compare in Figure
\ref{fig:pot13} the potential following from the analytical expression
(\ref{eq:phigen}) to its numerical counterpart. We do not 
display 1:1, 1:2 and 2:1 results since in these cases, Eq. (\ref{eq:phigen})
is obtained from an exact expansion and fully captures the
$r$-dependence of the potential. For the asymmetry 1:3,
Fig. \ref{fig:pot13} shows that the relatively simple form
(\ref{eq:phigen}) is very reliable. A similar agreement has been
found for all couples $z_-$:$z_+$ sampled, with the trend that 
the validity of (\ref{eq:phigen}) extends to larger distances
as $z_+/z_-$ is decreased. In this respect, the agreement
shown in Fig. \ref{fig:pot13} for which $z_+/z_-$ is quite high
(3), is one of the ``worst'' observed.

\begin{figure}[htb]
\vskip 4.5mm
\includegraphics[height=6cm,angle=0]{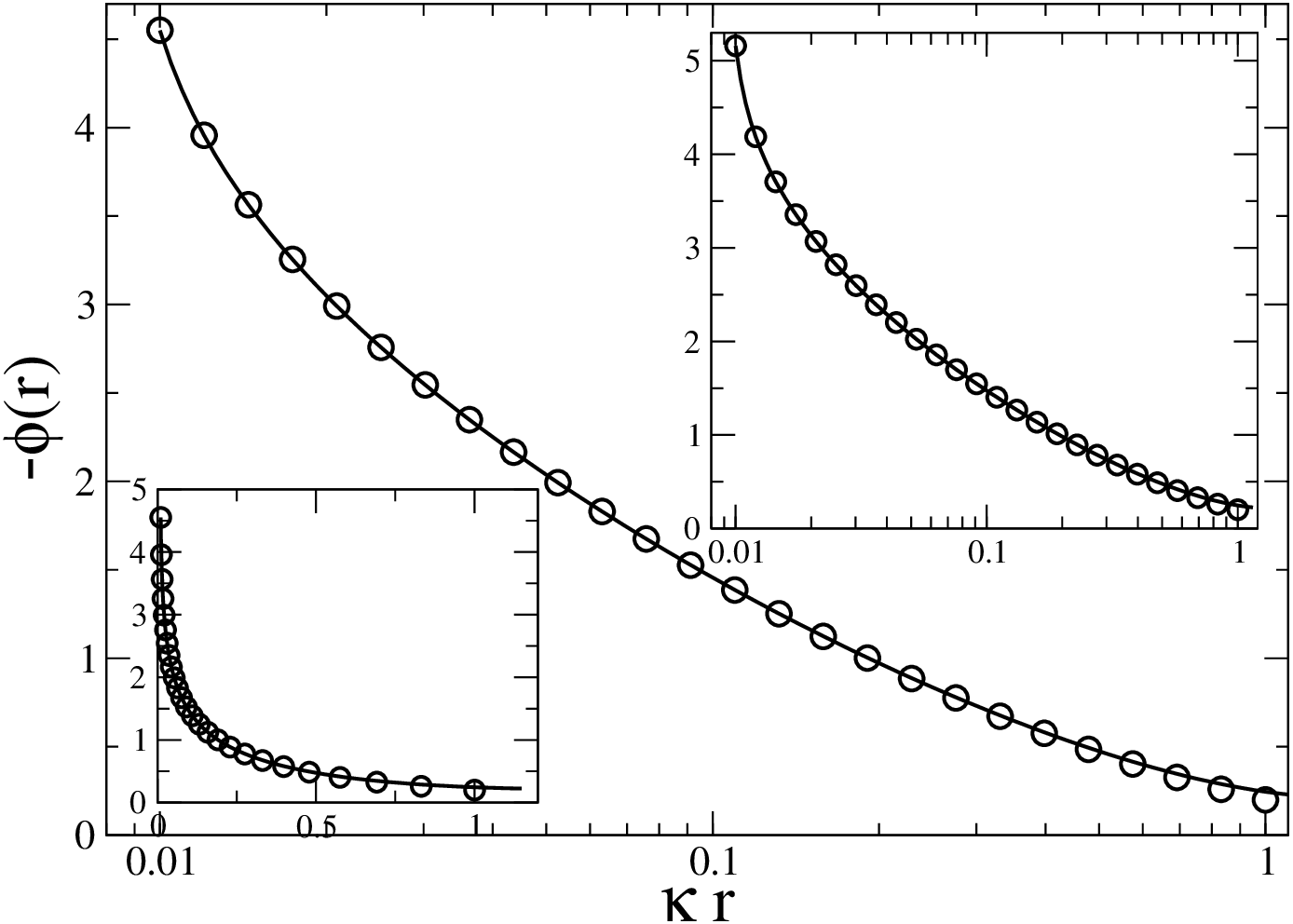}
\caption{Opposite of the 
electric potential versus radial distance in a 1:3 electrolyte
with $\kappa a =10^{-2}$.
The continuous curve shows the prediction of Eq. (\ref{eq:phigen})
with $\widetilde\mu$ given by $(\ref{eq:mutilde})$ ; 
the circles show the numerical solution of Eq. (\ref{eq:PB}).
The potential for $\xi=2.2$ is shown in the main graph
on a log-linear scale, and on a linear scale in the lower
inset. The upper inset is for $\xi=5$. 
\Label{fig:pot13}
}
\end{figure}

{\em Conclusion}. 
The poly-ion ion preferential interaction 
coefficient $\Gamma$  describes the exclusion of co-ions in the vicinity of
a polyelectrolyte in an aqueous solution. We have obtained
an accurate expression for $\Gamma$ in the regime of low
salt ($\kappa a <1$). The present results are particularly relevant for
highly charged poly-ions
($z_+ \xi >1$, that is beyond the classical Manning threshold
\cite{rque20}),
but are somewhat more general and hold in the range 
$\xi_c<\xi<1$, where $\xi$ stands for the line charge
per Bjerrum length and $\xi_c$ is a salt dependent threshold,
given by Eq. (\ref{eq:xicapp}). 
Our formulae have been shown to hold for arbitrary mixed salts
of the form $z_-$:$z_+$ (magnesium chloride, cobalt hexamine
etc). They have been derived from exact
expansions valid in 1:1,1:2 and 2:1 cases, from which a more
general conjecture has been inferred. The validity of this
conjecture, backed up by analytical arguments,
 has been extensively tested for various values
of $z_+/z_-$, poly-ion charge and salt content.
These tests have provided the numerical value of the 
constant $\cal C$ reported in Table \ref{table:1}, which only
depends of the ratio $z_+/z_-$.
As a byproduct of our
analysis, we have obtained a very accurate expression
for the electric potential in the vicinity of the charged rod
($r < \kappa^{-1}$).

It should be emphasized that the validity of our mean-field
description relying on the non-linear Poisson-Boltzmann equation
depends on the valency of counter-ions ($z_+$),
and to a lesser extent to the value of $z_-$ \cite{Levin,Grosberg}.
For the 1:1 case in a solvent like water at room temperature, micro-ionic correlations can be neglected up to a salt concentration of 0.1M \cite{Ni}. 
For $z_+\geq 2$ or in solvents
of lower dielectric permittivity,  they play a more
important role. Our results however provide mean-field benchmarks
from analytical expressions,
from which the effects of correlations may be assessed in cases
where they cannot be ignored (see e.g. \cite{Ni} for a detailed
discussion).

\begin{acknowledgments}
  This work was supported by a ECOS Nord/COLCIENCIAS action of French and
  Colombian cooperation. G.~T.~acknowledge partial financial support from
  Comit\'e de Investigaciones, Facultad de Ciencias, Universidad de los Andes.
  This work has been supported in part by the NSF PFC-sponsored Center for
  Theoretical Biological Physics (Grants No. PHY-0216576 and PHY-0225630).
\end{acknowledgments}

\begin{appendix}
\section{}
\Label{app:A}

In order to explicitly relate the preferential coefficient $\Gamma$ in (\ref{eq:Gamma})
to the electric potential, we follow a procedure similar to that which leads to an analytical 
solution in the cell model, without added salt \cite{Fuoss}. Implicit use will be made of the
boundary conditions associated to (\ref{eq:PB}). First, integrating 
Eq. (\ref{eq:PB}), one gets
\be
[r' \phi'(r')]_a^r \,=\, \frac{\kappa^2}{z_++z_-}  \int_a^r \left( e^{-z_+ \phi} - e^{z_-\phi} 
\right)\, r' dr' ,
\ee
where the notation $[F(r')]_a^r=F(r)-F(a)$ has been introduced.
Then, multiplying Eq. (\ref{eq:PB}) by $r^2 \phi'$ and integrating, we obtain
\begin{eqnarray}
\frac{z_++z_-}{2 \kappa^2}    \left[(r' \phi')^2 \right]_a^r &=& -\left[ r'^2 \frac{e^{-z_+ \phi}}{z_+}
+ r'^2  \frac{e^{z_-\phi}}{z_-}
\right]_a^r 
\nonumber\\
+&&\!\!\!\!\!\!\!\!\!
 \int_a^r  2r' \left( \frac{e^{-z_+ \phi}}{z_+}
+  \frac{e^{z_-\phi}}{z_-}
\right) dr' .
\end{eqnarray}
Combining both relations with adequate weights, in order to suppress 
the integral over counter-ion (+) density, we have
\begin{eqnarray}
\int_a^\infty r' (e^{z_- \phi}-1) dr'= \frac{ z_+ z_-}{\kappa^2} \left( \xi ^2 - \frac{2 \xi}{z_+}\right) 
\nonumber \\
- \frac{a^2}{2(z_++z_-)}  \left\{z_+\left(e^{z_-\phi_0} -1 \right) +z_- \left(e^{-z_+ \phi_0} -1\right)
\right\}
\label{eq:a3}
\end{eqnarray}
where $\phi_0=\phi(a)$ is the surface potential.
Equation (\ref{eq:a3}) will turn useful in the formulation 
of a general conjecture concerning the surface potential $\phi_0$, see 
Appendix \ref{app:B}. 
We also note that for the systems
under investigation here, the surface potential is quite high, and a very good approximation
to (\ref{eq:a3}) is
\be
\int_a^\infty r' (e^{z_- \phi}-1) dr' \simeq \frac{ z_+ z_-}{\kappa^2} \left( \xi ^2 - \frac{2 \xi}{z_+}\right) 
- \frac{a^2  z_- e^{-z_+ \phi_0} }{2(z_++z_-)} 
\label{eq:a4}
\ee

\section{}
\Label{app:B}

We start by analyzing a 1:1 electrolyte, for which it has been shown
\cite{McCoy,McCaskill} that the short distance behaviour reads
\be
e^{\phi/2} \,=\, \frac{\kappa r}{4 \mu}\, \sin\left[
2 \mu \log\left(\frac{\kappa r}{8}\right) - 2 \Psi(\mu)
\right] \,+\,
{\cal O}\left(\kappa r
\right)^4
\label{eq:grandpot}
\ee
where $\Psi$ denotes the argument of the Euler Gamma function 
$\Psi(x)=\hbox{arg}[\Gamma(i x)]$ \cite{McCoy,McCaskill}. 
In (\ref{eq:grandpot}), $\mu$ denotes the smallest positive root
of 
\be
\tan\left[
2 \mu \log(\kappa a/8) -2 \Psi(\mu) 
\right] \,=\, \frac{2 \mu}{\xi-1}.
\label{eq:mutan}
\ee
Expressions (\ref{eq:grandpot}) and (\ref{eq:mutan})
require that $\xi$ exceeds a salt dependent 
threshold [denoted $\xi_c$ below and given by Eq. (\ref{eq:xicapp})]
that is always smaller than 1 \cite{TT}.
They thus always hold for $\xi\geq1$ and in particular encompass the
interesting limiting case $\xi=1$,
which is sufficient for our purposes. 
For large $\xi$, we have proposed in 
\cite{TT} an approximation which amounts to linearizing the 
argument of the tangent in (\ref{eq:grandpot}) in the vicinity of $-\pi$,
and similarly linearizing  $\Psi$ to first order: 
$\Psi(x) \simeq -\pi/2 -\gamma x + {\cal O}(x^3)$ where $\gamma$ is 
the Euler constant, close to 0.577.
It turns out however that finding accurate expressions for
$\exp(-z_+\phi_0)$, which is useful for the computation of the
preferential interaction coefficient, requires to include the first non-linear
correction in the expansion of the tangent. 
After some algebra, we find~:
\begin{eqnarray}
\mu & \simeq &\frac{-\pi/2}{\log(\kappa a) +{\cal C} - (\xi-1)^{-1}} +
\nonumber\\
&&\frac{\pi^3}{6(\log(\kappa a) +{\cal C} - (\xi-1)^{-1})^4}
\nonumber\\
&&\times
\left[\frac{1}{(\xi-1)^3}+\frac{\psi^{(2)}(1)}{8}
\right]
\label{eq:approxmu}
\end{eqnarray}
where the constant ${\cal C}={\cal C}^{1:1}$ reads ${\cal C}^{1:1}=\gamma
-\log 8 \simeq -1.502$ and $\psi^{(2)}(1)=d^3\ln\Gamma(x)/dx^3|_{x=1}$. From
(\ref{eq:approxmu}) and (\ref{eq:grandpot}) where the sinus is expanded to
third order, we obtain \be (\kappa a)^2 e^{-\phi_0} \simeq 4 [(\xi-1)^2 +
\widetilde\mu^2]
\label{eq:phi0a}
\ee
where $\widetilde\mu$ is given by
\be
\widetilde\mu  \simeq \frac{-\pi}{\log(\kappa a) +{\cal C} - (z_+\xi-1)^{-1}}.
\label{eq:mutildeapp}
\ee
In writing (\ref{eq:mutildeapp}), we have introduced the change of 
variable $\widetilde\mu = 2 \mu$ \cite{rque1}. The reason is that similar changes 
for other electrolyte asymmetries allows to put the final result in a
``universal'' (electrolyte independent) form, see below. 
A similar reason holds for introducing $z_+$, here equal to 1,
in the denominator of (\ref{eq:mutildeapp}).

The functional proximity
between our expressions and those reported in \cite{Shkel} in the very same 
context is striking. We note however that our $\widetilde\mu$ (denoted
$\beta$ in \cite{Shkel}) involves a different constant ${\cal C}$.
More importantly, the functional form of (\ref{eq:grandpot}) differs from that
given in \cite{Shkel}. 
The comparison of the performances of our results with those
of \cite{Shkel} is addressed below, and is also discussed in the main 
text.

Performing a similar analysis as above in the 1:2 case where
$z_+=2$ and $z_-=1$, we obtain from the expressions derived
in \cite{TT}:
\be
(\kappa a)^2 e^{-z_+\phi_0} \simeq 3 [(\xi-1)^2 + \widetilde\mu^2]
\label{eq:phi0b}
\ee
and similarly, in the 2:1 case ($z_+=1$, $z_-=2$):
\be
(\kappa a)^2 e^{-z_+\phi_0} \simeq 6 [(\xi-1)^2 + \widetilde\mu^2].
\label{eq:phi0c}
\ee
In both cases, provided again that $\xi$ is not too low (see below)
$\widetilde\mu$ is given by (\ref{eq:mutildeapp}) \cite{rque2},
with however a different numerical value for $\cal C$
[${\cal C}^{1:2} =\gamma-(3\log3)/2-(\log2)/3\simeq -1.301 $
and ${\cal C}^{2:1} =\gamma-(3\log3)/2-\log2\simeq -1.763 $].

The similarity of expressions (\ref{eq:phi0a}), (\ref{eq:phi0b})
and (\ref{eq:phi0c}) leads to conjecture that this form holds
for any $z_-$:$z_+$ electrolyte :
\be
(\kappa a)^2 e^{-z_+\phi_0} \simeq {\cal A} [(z_+\xi-1)^2 + \widetilde\mu^2].
\label{eq:phi0d}
\ee
We then have to determine the prefactor $\cal A$ as a function 
of $z_+$ and $z_-$. To this end, we make use of the exact relation
(\ref{eq:a3}) [or equivalently (\ref{eq:a4})], where in the limit of
large $\xi$, the lhs is finite while the two terms on  the rhs diverge.
This yields the leading order behaviour :
\be
(\kappa a)^2 \exp(-z_+ \phi_0)  \stackrel{\xi\to\infty}{\sim} 2\,\frac{z_++z_-}{z_+} 
(z_+\xi-1)^2 .
\ee
It then follows  that ${\cal A} = 2 (z_++z_-)/z_+$ so that our general
expression (\ref{eq:phi0d}) takes the form:
\be
(\kappa a)^2 e^{-z_+ \phi_0} \, \simeq \, 2\,\frac{z_++z_-}{z_+} 
\left[(z_+\xi-1)^2  + \widetilde\mu^2\right].
\label{eq:phi0gen}
\ee
This expression holds regardless of the approximation used
for $\widetilde \mu$. If Eq. (\ref{eq:mutildeapp}) is used, then $z_+\xi$ should
not be too close to unity (see appendix \ref{app:C} for more general results
including the case $z_+\xi=1$).

In order to test the accuracy 
of (\ref{eq:phi0gen}) in conjunction with (\ref{eq:mutildeapp}), we have solved numerically Eq. (\ref{eq:PB})
for several values of $\kappa a<1$ and electrolyte asymmetry 
and checked that for several different values of $z_+ \xi>1$, 
the quantity 
\begin{eqnarray}
{\cal Q} &=& -\pi
\left[ (\kappa a)^2 e^{-z_+ \phi_0} \,\frac{z_+}{2(z_++z_-)} - 
(z_+\xi-1)^2
\right]^{-1/2}  \nonumber\\&&
-\log(\kappa a) + (z_+\xi-1)^{-1}
\label{eq:teststri}
\end{eqnarray}
is a constant  ${\cal C}$, which only depends on $z_+/z_-$ but not on salt and
$\xi$ [it should be borne in mind that Eq. (\ref{eq:mutildeapp}) 
is a small $\kappa a$
and large $\xi$ expansion, which becomes increasingly incorrect
as $\kappa a$ is increased and/or $\xi$ lowered].
This is quite a stringent test (since the two terms on the rhs of
(\ref{eq:teststri}) are large and close] which requires high numerical 
accuracy. This is achieved following the procedure outlined in 
\cite{Lang}. In doing so, we confirm the validity of
(\ref{eq:phi0gen}) and collect the values of $\cal C$ given in Table
\ref{table:1}. In the 1:1 case, we predict that ${\cal C} = \gamma-\log8
\simeq -1.507$, in excellent agreement with the numerical
data of Figure \ref{fig:Q1}.
On the other hand, the prediction of Ref. \cite{Shkel} 
that  $\cal Q$ reaches a constant close to  -1.90 (shown by the horizontal
dashed line in Fig \ref{fig:Q1}) is incorrect. Figure \ref{fig:Q1} shows
that the quality of expression (\ref{eq:phi0gen}) deteriorates when 
$\kappa a$ increases, as expected. 
It is noteworthy however that for $\kappa a =10^{-1}$,
its accuracy is excellent whenever $\xi>2$. The inset 
of Fig. \ref{fig:Q1} shows the validity of (\ref{eq:phi0gen}) 
for a 3:1 electrolyte. When $z_+\xi$ is close to 1, 
Eq. (\ref{eq:mutildeapp}) becomes an irrelevant approximation 
to the solution of (\ref{eq:mutan}), and can therefore not be
inserted into the general formula (\ref{eq:phi0gen}).
This explains the large deviations between $\cal Q$
and the asymptotic value ${\cal C}$ observed in Fig. \ref{fig:Q1}
for the lower values of $\xi $ reported.
We come back to this point in Appendix \ref{app:C}.

\begin{figure}[htb]
\vskip 4.5mm
\includegraphics[height=6cm,angle=0]{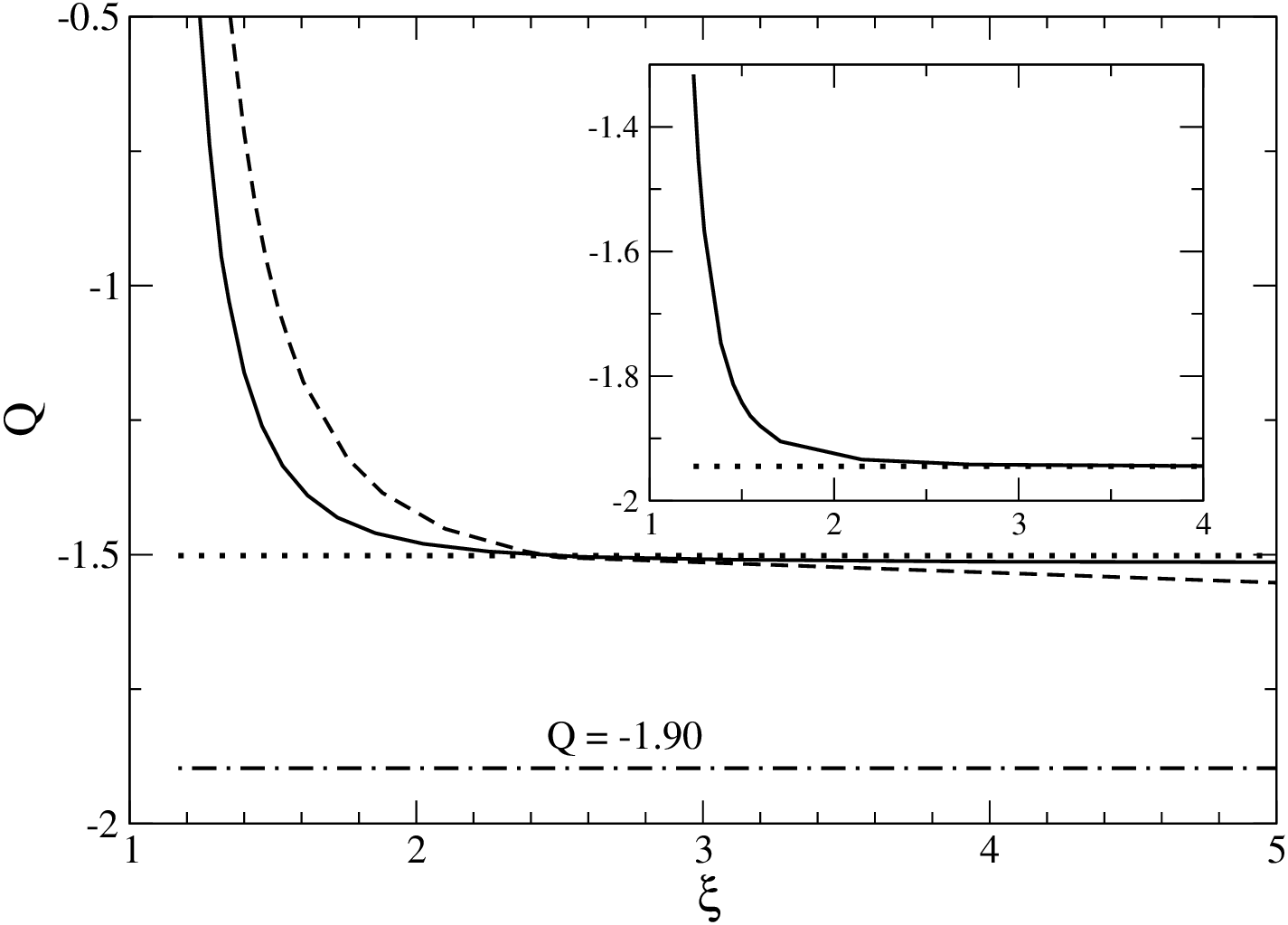}
\caption{Plot of the quantity $\cal Q$ defined in (\ref{eq:teststri}) 
versus line charge $\xi$ for a 1:1 electrolyte at $\kappa a =10^{-3}$ 
(continuous curve) and $\kappa a=10^{-1}$ (dashed curve).  
The value reached at large $\xi$ is compared to 
the prediction of \cite{Shkel}  ${\cal Q} \to e^\gamma + \log 2-\gamma\simeq 
-1.90$ (horizontal dashed-dotted line) 
whereas Eqs. (\ref{eq:phi0gen})
and (\ref{eq:mutildeapp}) imply ${\cal Q} \to \gamma-\log8 \simeq
-1.50$, shown by the horizontal
dotted line. The inset shows the same quantity for a 3:1 electrolyte at 
$\kappa a=10^{-5}$ [such a very low value is required to determine
precisely the value of the asymptotic constant $\cal C$,
that can subsequently be used at experimentally relevant (higher)
salt concentrations]. Here, we obtain ${\cal Q}  \to -1.94$ (dotted line)
which is the value reported for $\cal C$ in Table \ref{table:1}.
\Label{fig:Q1}
}
\end{figure}

The present results  hold for $z_+\xi>1+ {\cal O}(1/|\log \kappa a|)$. 
In this regime, our analysis shows that Eq. (\ref{eq:phi0gen})
[with $\widetilde\mu$ given by (\ref{eq:mutildeapp})] is correct up to
order $1/\log^4(\kappa a)$ for any ($z_-,z_+$). On the other hand 
the results of \cite{Shkel} ,valid in the 1:1 case, appear to be correct
to order $1/\log^2(\kappa a)$.
In addition, our expression for the surface potential 
may be generalized to a broader range of $\xi$ values and 
an expression for the short distance dependence of the electric potential
may also be provided. This is the purpose of appendix 
\ref{app:C}.

\section{}
\Label{app:C}

In Appendix \ref{app:B}, the ``universal'' results valid for all 
($z_+,z_-$) have been unveiled partly by a change of variable 
$\mu \to \widetilde\mu$ from existing expressions \cite{TT}. In light of these
results, and of their accuracy (assessed in particular by the precision 
reached for the preferential interaction coefficient), it is tempting to 
go further without invoking approximations of (\ref{eq:mutan}), or related 
expressions for other asymmetries than 1:1. Inspection of the 
results given in \cite{TT} for the 1:1, 1:2 and 2:1 cases lead, with again the help of (\ref{eq:a4}), to the conjecture that 
\be
e^{z_+\phi/2} \,\simeq\, 
\frac{-\kappa r}{ \widetilde\mu}\, \sqrt{\frac{z_+}{2(z_++z_-)}} \,\sin\left[
\widetilde \mu \log(\kappa r) +\widetilde\mu\,{\cal C}
\right]
\label{eq:phigen}
\ee
with 
\be
\tan\left[\,
\widetilde\mu \log(\kappa a) \,+\, \widetilde\mu \, {\cal C}
\,\right] \,=\, \frac{\widetilde\mu}{z_+\xi-1}.
\label{eq:mutantilde}
\ee We emphasize that (\ref{eq:phigen}), much as (\ref{eq:grandpot}), is a
short distance expansion and typically holds for $\kappa r<1$ (hence the
requirement that $\kappa a<1$).  In appendix~\ref{app:D} we give further
analytical support for conjecture~(\ref{eq:phigen}).  A typical plot showing
the accuracy of (\ref{eq:phigen}) is provided in the main text (Fig.
\ref{fig:pot13}). For $\kappa r <0.1$, the agreement 
with the exact result is 
better than 0.1\%, 
and becomes progressively worse at higher distances
(20\% 
disagreement at $\kappa r=1$).

From (\ref{eq:phigen}), it follows that the integrated charge $q(r)$
in a cylinder of radius $r$ [that is $q(r)=-r \phi'(r)/2$] reads 
\be
z_+ q(r) \,=\, -1 + \widetilde\mu \,\tan \left[\widetilde\mu \, \log 
\left(\frac{r}{R_M} \right)   \right]
\ee
where the so-called Manning radius \cite{Gueron,OS,TT} is given by 
\be
\kappa R_M = \exp\left(-{\cal C} -\frac{\pi}{2\widetilde\mu}
\right).
\ee
The Manning radius is a convenient measure of the counterion 
condensate thickness. It is the point $r$ where not only $z_+ q(r) =1$
but also where $q(r)$ versus $\log r$ exhibits an inflection point
\cite{Deserno}. For high enough $\xi$, the logarithmic dependence 
of $1/\widetilde\mu$ with salt [see (\ref{eq:mutildeapp})] is such that
$R_M \propto \kappa^{-1/2}$.

The two relations (\ref{eq:phigen}) and (\ref{eq:mutantilde})
encompass those given in Appendix \ref{app:B} and allow to 
investigate the regime $z_+\xi_c<z_+\xi$, and in particular 
the case $z_+\xi=1$, the so-called Manning threshold \cite{Manning}.
However, (\ref{eq:phigen}) and (\ref{eq:mutantilde}) are not valid 
for $\xi<\xi_c$, with 
\be
z_+ \xi_c \,=\, 1 + \frac{1}{\log\kappa a+{\cal C}}.
\label{eq:xicapp}
\ee
Note that $\xi_c<1$, since the constant $\cal C$ is negative and that 
salt should fulfill $\kappa a<1$. For $\kappa a=10^{-2}$
and $z_+/z_-=1$, we obtain $\xi_c \simeq 0.836$. This is
precisely the point where $-\Gamma =1$ in the inset of Fig.
\ref{fig:Gamkap0.01}. This inset also shows that the value
of $\Gamma$ resulting from the use of the solution of (\ref{eq:mutantilde})
is remarkably accurate.

At this point, it seems useful to investigate  the Manning threshold case
$z_+ \xi =1$ (which corresponds to the onset of counterion  condensation 
when $\kappa a \to 0$ \cite{Manning,TT,Deserno}). It is readily seen that 
the solution of (\ref{eq:mutantilde}) reads
\be
\widetilde\mu ~~\stackrel{z_+\xi=1}{=} ~~
\frac{-\pi/2}{\log(\kappa a) +{\cal C}},
\label{eq:muxi1app}
\ee
which should be inserted in (\ref{eq:phigen}) to obtain the potential 
profile, or in (\ref{eq:Gammaapprox}) to get the interaction coefficient.

\section{}
\Label{app:D}

In this appendix we give further support for the conjecture~(\ref{eq:phigen})
which gives the short-distance expansion of the electric potential. Let us
suppose initially that the charge is below the Manning threshold $\xi<\xi_c$.
It is straightforward to verify that Poisson--Boltzmann equation~(\ref{eq:PB})
admits solutions which behave as $\phi(r)=-2A\ln(\kappa r) + \ln B + o(1)$ for
$\kappa r \ll 1$. Injecting this expansion into
equation~(\ref{eq:PB}) allows us to compute higher order terms. To study the
regime beyond the Manning threshold, we compute all higher order terms of the
form $r^{2 n (1+z_{+}A)}$ (for a negatively charged macroion) and $r^{2n
  (1-z_{-}A)}$ (for a positively charged macroion), with $n$ a positive
integer. These terms turn out to present themselves as the series expansion of
the logarithm, thus resumming them we obtain
\begin{eqnarray}
  \label{eq:phi-expans}
  \phi(r)&=&-2A\ln(\kappa r)+ \ln B\\
  && 
  \nonumber
  +\frac{2}{z_+}\ln\left[1-\frac{z_{+} B^{-z_{+}}\, (\kappa r)^{2(1+z_{+}
  A)}}{8(z_{+}+z_{-})(1+z_{+}A)^2}
    \right]\\
    \nonumber
    &&-\frac{2}{z_{-}}\ln\left[1-\frac{z_{-} B^{z_{-}}\, (\kappa r)^{2(
          1-z_{-}A)}}{8(z_{+}+z_{-})(1-z_{-}A)^2}
    \right]+\cdots
\end{eqnarray}
The dots represent terms of order $r^{2n(1+z_{+}A)+2m(1-z_{-}A)}$ with $n$ and
$m$ two nonzero positive integers. When the Manning threshold is approached,
$z_{+}A+1=0$ for negatively charged macroion, the terms $r^{2 n (1+z_{+}A)}$
(second line of Eq.~(\ref{eq:phi-expans})) become of order one, but the rest
of the terms (third line of Eq.~(\ref{eq:phi-expans}) and dots) remain higher
order: a change in the small distance behavior of $\phi$ occurs.  A similar
situation is reached for $1-z_{-}A=0$ which is the Manning threshold for a
positively charged macroion.

$A$ and $B$ in the previous equations are constants of integration, which
should be determined with the boundary conditions $r \phi'(r) = 2\xi$ at the
polyion radius ($r=a$) and $\phi \to 0$ for $r\to \infty$. Thus to proceed
further, we have to connect the long and the short distance behavior of
$\phi$. This connection problem has been only solved in the cases 1:1, 1:2 and
2:1 in Refs.~\cite{McCoy,TracyWidom-Toda-asympt}. In particular, once $A$ has
been chosen (notice that for $a=0$, $A=-\xi$), $B$ should be one and only one
function of $A$ in order to satisfy $\phi \to 0$ for $r\to \infty$. The
results from~\cite{McCoy,TracyWidom-Toda-asympt} show that
\begin{eqnarray}
  \label{eq:B}
  B&=&2^{6A}\gamma\left((1+A)/2\right)^2\quad(1:1)\\
  B&=&3^{3A}2^{2A}\gamma\left(2(1+A)/3\right)\gamma\left((1+A)/3\right)
  \quad(1:2)\nonumber\\
  B&=&3^{3A}2^{2A}\gamma\left((1+2A)/3\right)\gamma\left((2+A)/3\right)
  \quad(2:1)\nonumber
\end{eqnarray}
where $\gamma(x)=\Gamma(x)/\Gamma(1-x)$. $B$ turns out to have some
interesting properties in the cases 1:1, 1:2 and 2:1, where its exact
expression~(\ref{eq:B}) is known. Namely, at the
Manning threshold $1+z_{+} A=0$, 
\begin{equation}
  \lim_{A\to-1/z_{+}}
  \frac{z_{+} B^{-z_{+}}}{8(z_{+}+z_{-})(1+z_{+}A)^2}=1
\end{equation}
Furthermore if we put $1+z_{+}A=i\widetilde\mu$, and define
\begin{equation}
  \label{eq:defPsi}
  e^{2i\Psi(\widetilde\mu)}= \frac{z_{+} B^{-z_{+}}}{
    8(z_{+}+z_{-})(1+z_{+}A)^2}
\end{equation}
then for $\widetilde\mu\in\mathbb{R}$, $\Psi(\widetilde\mu)\in\mathbb{R}$ is a
\textit{real} function of $\widetilde\mu$, with $\Psi(0)=0$.

Let us now study the regime beyond the Manning threshold for a negatively
charged macroion. From Eq.~(\ref{eq:phi-expans}) we can write
\begin{equation}
  \label{eq:expphi-start}
  e^{z_{+}\phi(r)/2}\sim
  (\kappa r)^{-z_{+}A}B^{z_{+}/2}
  \left(1-\frac{z_{+} B^{-z_{+}}\, (\kappa r)^{2(1+z_{+}
        A)}}{8(z_{+}+z_{-})(1+z_{+}A)^2}
  \right)
\end{equation}
neglecting terms of higher order when $z_{+} A$ is close to $-1$. 

Let us conjecture that the properties of $B$ as a function of $A$ presented
above hold in the general case $z_{-}:z_{+}$. Then using the parameter
$\widetilde\mu$ defined above we find after some simple algebra
\begin{eqnarray}
  \label{eq:phigeneral}
  e^{z_{+}\phi(r)/2}
  &=&
  \frac{-\kappa r}{ \widetilde\mu}\, \sqrt{\frac{z_+}{2(z_++z_-)}} \,\sin\left[
    \widetilde \mu \log(\kappa r) +\Psi(\widetilde\mu)
  \right]  \nonumber\\
  &&
  +O(r^{3+2z_{-}/z_{+}})
\end{eqnarray}
Recalling that $|\widetilde\mu|\ll 1$ we can approximate
$\Psi(\widetilde\mu)\simeq \widetilde\mu {\cal C} $, where ${\cal
  C}=\Psi'(0)$.  Replacing this approximation into~(\ref{eq:phigeneral}) and
imposing the boundary condition $a \phi'(a) = 2\xi$ leads to~(\ref{eq:phigen})
and~(\ref{eq:mutantilde}). Numerical values obtained for the constants ${\cal
  C}$ are reported in Table~\ref{table:1}, for different charge asymmetries
$z_{-}:z_{+}$. The previous analysis shows that analytical predictions for
${\cal C}$ could be made if the connection problem is solved and the
equivalent of expressions~(\ref{eq:B}) are found for the general case
$z_{-}:z_{+}$.

\end{appendix}



\begin{thebibliography}{99}
\bibitem{rque10}
In the 1:1 case, our definition differs
from the more standard one as found e.g. in \cite{Shkel}
by a factor $4\xi$. The reason for doing so is that this
allows easier comparison of the salt dependence of $\Gamma$ 
for different values of the poly-ion charge.
\bibitem{rque20}
We emphasize that accurate results for $\Gamma$, 
$\phi$ etc may be obtained for $\xi <\xi_c$ from the results
given in \cite{TT}. We did not investigate this regime here,
since it is of little relevance for nucleic acids.
\bibitem{rque1}
It then appears that the expression given for $\widetilde\mu=2\mu$
in (\ref{eq:mutildeapp}) corresponds to the dominant
term only in (\ref{eq:approxmu}) (the first one on the rhs).
\bibitem{rque2}
Compared to the expressions given in \cite{TT} where a parameter 
$\mu$ plays a key role, the corresponding change of variables
should be performed: $\widetilde\mu=3\mu$ (1:2 case) and
$\widetilde\mu=3\mu/2$ for 2:1 electrolytes.
\end{thebibliography}
\end{document}